# Optical control of coherent and squeezed phonons: major differences and similarities

**Misochko O.V.**

*Institute of Solid State Physics, Russian Academy of Sciences*

*142432 Chernogolovka, Russia*

*misochko@issp.ac.ru*

Coherent and squeezed phonon oscillations can be excited in solids impulsively by a single femtosecond pulse whose duration is shorter that a phonon period. By applying the second ultrafast pump pulse these oscillations can be significantly, but differently modified.

Progress in femtosecond lasers and ultrafast spectroscopy has enabled us to generate and observe the coherent and squeezed lattice oscillations in which atomic motions are excited with light through a nonlinear process and the resulting dynamics is directly observed in the time domain [1]. The process of generation of coherent and squeezed phonons by using ultrashort light pulses has provided a unique tool for time-domain studies of optical and acoustic modes and their coupling to electronic and other excitations in solids. To date such phonons have been observed in time-resolved optical pump-probe experiments as periodic modulations of reflectivity or transmission with increasing attention being turned towards not only observation but also understanding of physics which occurs at this short time scale. The coherent control of these phonons is one of the key issues in femtosecond technology. Such control is enabled by the full knowledge of the dynamics of crystal lattice concerning its amplitude and phase. It is due to the existence of a well-defined phase, phonons generated by an ultrafast pulse can be easily manipulated. Thus, providing well-defined quanta of energy in a well-defined temporal sequence, the lattice can be driven into an artificial quantum-mechanical state, which cannot be achieved by other means of optical excitation. Although interference due to which such control is achieved is intrinsically a classical phenomenon, the superposition principle which underlies it is also at the heart of quantum mechanics. Indeed, in some interference experiments we encounter the idea of quantum entanglement, which became clear after the famous paper by Einstein, Podolsky and Rosen singled out some startling features of the quantum mechanics. Schrödinger emphasized that these unusual features are due to the existence of what he called "entangled states", which are two-particle states that cannot be factored into products of two single-particle states in any representation. "Entanglement" is just Schrödinger's name for superposition in a composite system. Therefore, in this work we have attempted to compare coherent control of one- (coherent) and two-phonon (squeezed) lattice states created by ultrafast laser pulse. In particular, we report on the control of coherent $A_{1g}$ optical phonons in Bi and of squeezed acoustical biphonons in ZnTe using two pump pulse technique.

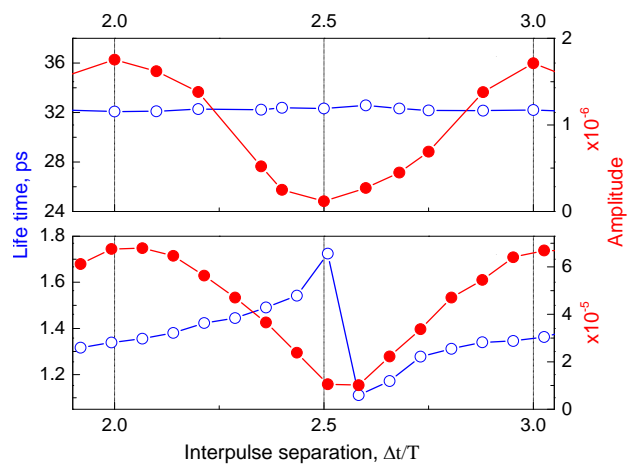

Fig.1. Lifetime (open symbols, left scale) and amplitude (filled symbols, right scale) as a function of control parameter for coherent phonons in Bi (top) and squeezed biphonons in ZnTe (bottom).

To study the optical control over coherent phonons, we performed two pump–single probe reflectivity experiments on Bi crystal at helium temperature. It was observed that the reflectivity oscillations due to coherent $A_{1g}$ phonons can be modified by a second pump pulse [2]. Namely, a time delay of nT between the two pump pulses, where n is an integer and T=331fs is the phonon period, resulted in enhancement while a time delay of (n+1/2)T resulted in cancellation of the oscillations, see Fig.1. Even though there was a singularity in the phase behavior at (n+1/2)T, lifetime of the coherent phonons was unaffected by the second pump pulse, as shown in Fig.1. The latter is compatible with the results of passive control [4] at low excitation strength, where the coherent amplitude grows linearly, while the lifetime remains unchanged for the pump fluences not exceeding 100μJ/cm$^2$. Given that the $A_{1g}$ coherent phonons in bismuth are driven by photoexcited carriers (displacive mechanism), the process of coherent control can be thought of in the following way: The first pump pulse creates a new potential surface on which the atoms move. Initially displaced from the newly established equilibrium configuration, the lattice achieves this configuration in approximately one quarter of a phonon period, but the atoms have momentum at that point. When the atoms reach the classical turning point of their motion, a second pump pulse can excite the precise density of carriers to shift the equilibrium position to the current position of the atoms, stopping the oscillatory motion. Because photoexcitation of additional electrons can displace the potential only in one direction, the vibrations can only be stopped at one of the classical turning points.

Having discussed the first-order control of lattice displacements, in which we manipulate interatomic separations, next we turned to second-order control. When ultrafast laser pulse strikes a crystal with van Hove singularity in the phonon density of states, it creates a pair of correlated acoustic phonons running in opposite directions. As a result, the atomic fluctuations in either position or momentum become squeezed below the vacuum level for a quarter of the oscillatory cycle. Because of a correlated behaviour (annihilation operator of one half of the pair is correlated/anticorrelated to creation/annihilation operator of the other half), both halves interfere destructively resulting in two-phonon coherence and a zero net population for the one-phonon mode. Thus, following the impulsive optical excitation, which macroscopically excites lattice vibrations, the transmission in ZnTe starts to oscillate [3]. The oscillation lifetime and frequency at room temperature are 1.4 ps and 3.67 THz, the latter being close to but slightly higher than that of 2TA(X) overtone. Coherent amplification and suppression of the two-phonon oscillations is achieved by two-pump, single-probe technique which, in this case, is an acoustical analogue of two-photon interference for parametrically downconverted photons. Here instead of overlapping pair-correlated photons, we overlap two ensembles of pair-correlated phonons created at different times. The whole control process can be described as the sum of two ensembles of phonons whose motion was initiated at different times and that now interfere. The relative timing of pump pulses determines whether the oscillations add constructively or destructively. As shown in Fig.1, depending on the interpulse separation, the lifetime, as well as the amplitude of two-phonon oscillations, is modulated. The modulation of lifetime, which was absent for coherent phonons, suggests that we are dealing with quantum interference as for its classical counterpart the change of lifetime is impossible: classical coherent states are always transformed into different coherent states. It should be noted that the change of amplitude can be either correlated or anti-correlated to that of life time. The two-phonon amplitude scales with the real part of squeeze parameter, whereas the oscillation lifetime measures how long the constituents of two-phonon state are correlated, or entangled: the longer the time, the stronger the entanglement [3]. Naively, one might think that a larger squeezing results in a stronger entanglement since quantum entanglement is closely related to quantum squeezing. However, as follows from the data presented in Fig.1, the extreme life times come about near *minimum* of the amplitudes, or when the sign of resulting biphonon amplitude changes from positive to negative. Moreover, the decrease in lifetime occurs over a shorter time scale as compared to its increase. This suggests that the relationship between phonon squeezing and entanglement may be more complicated because these effects are of different orders. The squeezing is the second order effect relying on the amplitude-amplitude correlations controlled by one-phonon interference. The entanglement, in contrast, belongs to a class of the forth order effects governed by intensity-intensity correlations (the same class as bunching/anibunching phenomena dictated by field statistics).

In conclusion, we have demonstrated that coherent and squeezed phonon oscillations can be excited in solids impulsively by a single femtosecond pulse whose duration is shorter that a phonon period. By applying the second ultrafast pump pulse these oscillations can be significantly, but differently modified. Thus, the ultrafast laser pulses provide a flexible and powerful tool not only to create and observe but also to control phonon amplitude, squeezing and entanglement. First-order control allows manipulating interatomic separations, whereas second-order control provides the possibility to modify uncertainty contours for the lattice atoms keeping their separations intact.

1. K. Ishioka, and O.V. Misochko, in Progress in Ultrafast Intense Laser Science **V**, Eds. K.Yamanouchi, A. Giullietti, and K. Ledingham, 23-64, Springer Series in Chemical Physics, Berlin (2010).
2. O.V. Misochko, and M.V. Lebedev, JETP Letters **90**, 284 (2009).
3. O.V. Misochko, J. Hu, and K.G. Nakamura, Physics Letters A 375, 4141 (2011).
4. O.V. Misochko, and M.V. Lebedev, JETP 109, 805 (2009).